# MIS Quarterly

# FROM REPRESENTATION TO MEDIATION: A NEW AGENDA FOR CONCEPTUAL MODELING RESEARCH IN A DIGITAL WORLD[1]

**Jan Recker**
Faculty of Management, Economics, and Social Sciences, University of Cologne,
Albertus-Magnus-Platz, 50923 Köln, GERMANY  {jan.recker@wiso.uni-koeln.de}

**Roman Lukyanenko**
Department of Information Technologies, HEC Montréal, 3000 Chemin de la Côte-Sainte-Catherine,
Montréal, QC,  H3T 2A7  CANADA  {roman.lukyanenko@hec.ca}

**Mohammad Jabbari**
School of Information Systems, Queensland University of Technology, 2 George Street,
Brisbane,  QLD 4000  AUSTRALIA  {m.jabbarisabegh@qut.edu.au}

**Binny M. Samuel**
Lindner College of Business, University of Cincinnati, 2906 Woodside Drive,
Cincinnati, OH  45221  U.S.A.  {samuelby@uc.edu}

**Arturo Castellanos**
Zicklin School of Business, Baruch College, City University of New York, 55 Lexington Avenue,
New York, NY  10010  U.S.A.  {Arturo.Castellanos@baruch.cuny.edu}

*The role of information systems (IS) as representations of real-world systems is changing in an increasingly digitalized world, suggesting that conceptual modeling is losing its relevance to the IS field. We argue the opposite: Conceptual modeling research is more relevant to the IS field than ever, but it requires an update with new theory. We develop a new theoretical framework of conceptual modeling that delivers a fundamental shift in the assumptions that govern research in this area. This move can make traditional knowledge about conceptual modeling scripts as mediators between physical and digital realities. Our framework draws attention to the role of conceptual modeling scripts as mediators between physical and digital realities. We identify new research questions about grammars, methods, scripts, agents, and contexts that are situated in intertwined physical and digital realities. We discuss several implications for conceptual modeling scholarship that relate to the necessity of developing new methods and grammars for conceptual modeling, broadening the methodological array of conceptual modeling scholarship, and considering new dependent variables.*

**Keywords**:  Conceptual modeling, mediation, digital technology, ontological reversal, digital objects, representation

# Introduction ███████████

Nick Carr's (2003) article "IT Doesn't Matter" appears as a remnant from long-forgotten times. The information systems (IS) world has changed considerably since those words were written. Technological developments in software and hardware have made it possible to infuse digital technologies into a wide range of traditional economic goods, from airplanes to cars, from kitchen scales to sound systems and toothbrushes (Faulkner and Runde 2019; Yoo 2010). Digital technologies have changed the nature and structure of economic goods (Porter and Heppelmann 2014), enabled radically new business processes such as crowd-based innovations (e.g., Bayus 2013), spawned novel business models such as data-driven businesses (e.g., Parmar et al. 2014), and even transformed entire industries, including transport, hospitality, and finance (e.g., Iansiti and Lakhani 2014; Porter and Heppelmann 2014). In a world that abounds with digital technologies, modern IS no longer only represent reality (Burton-Jones et al. 2017; Recker et al. 2019) but increasingly shape it (Alaimo and Kallinikos 2017; Baskerville et al. 2020).

This fundamental shift in the role of IS has profound implications for conceptual modeling (CM), that is, the development and use of representations to capture the features of a real-world domain an IS is intended to support (Mylopoulos 1992; Wand and Weber 2002). CM has been a core IS research area since the field's formative years (Bubenko 1979; Weber 1987) and is still being actively pursued (e.g., Burton-Jones et al. 2017; Rai 2017).

CM research has developed important insights based on several assumptions that have guided it to this point, but these assumptions are challenged by the current IS landscape:

- Movements like agile development (Fowler and Highsmith 2001) and DevOps (Wiedemann et al. 2019) alter modeling and documentation practices during systems analysis and design, which has traditionally been a key area where CM is used.

- Technological developments like NoSQL databases, machine learning, and business analytics challenge the form, function, and utility of relational databases (Storey and Song 2017), which have traditionally been a key IS component designed through CM.

- The ongoing infusion of digital technologies into economic goods and everyday artifacts blurs the boundaries among the surface, physical, and deep structures of IS. Traditionally, a main focus of CM was on the deep structure meaning.

- Actions and decisions taken in a digital reality increasingly influence those in a physical reality (Baskerville et al. 2020). CM has traditionally provided guidance for the development of digital representations of physical reality alone.

- Collective action movements like open source development (Bagozzi and Dholakia 2006), citizen science (Levy and Germonprez 2017), and crowdsourcing (Majchrzak and Markus 2013) increasingly involve nonspecialist users in IS development. Traditionally, only trained IS professionals engaged in such activities.

The time is ripe to update CM theory because the assumptions and approaches of CM scholarship increasingly underrepresent and constrain CM's potential to support the ongoing digitalization of reality. Therefore, we offer a new framework for CM scholarship that is fit for the burgeoning digital world.

To develop this framework, we engage in a dialectical argument that involves looking back and forward. In looking back, we identify and challenge historically grown field assumptions, that is, broad sets of beliefs about central aspects of a subject (Alvesson and Sandberg 2011), in the CM literature. In looking forward, we identify changes in the IS landscape that are relevant to CM by drawing on the emergent body of theoretical work around the nature of digital objects (Alaimo and Kallinikos 2017; Faulkner and Runde 2019) and their profound implications for how IS scholarship should be situated in a digital world (Baskerville et al. 2020; Yoo 2010). Then we develop a new conceptualization of CM that includes updated assumptions about the core constructs of CM scholarship, that is, scripts, grammars, methods, and context (Wand and Weber 2002). The novel idea of our framework is that the role of CM in a digital world is to assist *both* representation and shaping. We conceptualize this new role for CM as the task of *mediating* the transition from—and between—the states of physical and digital realities.[2]

Our new conceptualization connects past CM research to the future and provides a new, broad theoretical platform for CM research that illuminates new and unexplored research ques-tions. We outline selected research areas that lie at the intersection of physical to digital and digital to physical realities and between digital realities. We close by discussing broad implications for how CM scholarship could be carried out in the future.

---

[2]We use the terms *physical reality* and *digital reality* to indicate the difference between *lived* versus *computed* human experiences. Like Burton-Jones et al. (2017, p. A2), we conceptualize physical reality as "the aggregation of constituent things and their properties that exist in the real world, as perceived by someone or some group," which includes social as well as material things. In this understanding, the term *physical reality* includes not only tangible, material objects (e.g., buildings, cars, products, technologies, people) but also social constructions (e.g., organizations, processes, contracts, relationships) that feature in lived experiences. This understanding is broader than Baskerville et al.'s (2020) use of the term *physical*, which we interpret as being limited to material things.

## Background on Conceptual Modeling ■

Since the 1970s, IS professionals, such as business and process analysts, systems designers, and software developers, have used semi-formal, often graphic, representations to analyze IS, design IS, or visualize datasets. These representations, which are commonly called *conceptual models*, describe an individual's or group's understanding of a real-world domain and the features or phenomena in that domain (Kung and Sølvberg 1986; Mylopoulos 1992; Wand and Weber 2002).

Conceptual models are *scripts* (products of the CM process) developed by using *grammars* (sets of constructs and the rules by which to combine them) and guided by a *method* (procedures by which a grammar can be used) within an organizational *context* (the setting in which scripts are developed and used) (Wand and Weber 2002). CM scripts can be used to represent surface structures (facilities in an IS that allow users to interact with the IS, such as a graphic user interface) or physical structures (the underlying technology that is used to operate the IS, such as a network infrastructure) (Weber 1997, pp. 78-80). However, the main focus of CM scripts is the *deep structure* of an IS, that is, its characteristics that manifest stakeholders' perception of the meaning of the real-world phenomena it is intended to represent (Wand and Weber 1995, pp. 205-207).

During the early years of CM, generally the 1970s to the 1990s, many studies focused on developing modeling grammars for systems analysis and database design (Chen 1976). In the mid-1990s, emerging practices such as business process reengineering and object-oriented analysis and design created another wave of CM research in which new methods and grammars were developed, including those that focused on object or process modeling (Scheer 1994; Vessey and Conger 1994).[3] Also at that time, criticisms arose that suggested that the fundamental concepts and underlying methods of CM research were not clearly defined (Batra and Marakas 1995; Wand and Weber 1995). In response to these arguments, researchers developed and proposed frameworks to provide a more structured approach for how CM research could be pursued (e.g., Hirschheim et al. 1995; Lindland et al. 1994; Topi and Ramesh 2002; Wand and Weber 2002). After publication of these frameworks, another steady wave of CM research emerged that largely focused on evaluating the capabilities of existing CM grammars and scripts, rather than building new modeling grammars or methods (Burton-Jones et al. 2009; Recker et al. 2019).

More recently, two facets of CM scholarship have emerged: On one hand, CM research continues to be published in the field's top journals, including *MIS Quarterly,* addressing topics such as the CM of user-generated content for citizen science (Lukyanenko, Parsons et al. 2019), various theoretical bases on which to improve the quality of CM (Clarke et al. 2016), new experimental findings on CM use (Samuel et al. 2018), and new methodological procedures for CM research (Bera et al. 2019; Lukyanenko, Parsons, et al. 2019). On the other hand, an increasing number of critical accounts that suggest that CM may be becoming "obsolete" (Lukyanenko and Parsons 2013) or that fundamental CM ideas may be "dying" (Atzeni et al. 2013; Fowler 2001) have been published. As Wand and Weber (2017, p. 1) wrote,

> the topic of conceptual modeling lacks the appeal of research on emerging technologies (because it is deemed to be an old-technology problem) …. Thus, young scholars, in particular, have shied away from the topic.

We interpret this facet of CM scholarship as signaling that the CM field is at a tipping point, and it is time to move the field forward.

## Revisiting the Assumptions Around Conceptual Modeling in a Digital World ■

Even though IS scholars have produced quality research on CM, the topic has remained a niche area within the IS community (Burton-Jones et al. 2017; Wand and Weber 2017), perhaps because of the somewhat narrow agenda CM researchers have pursued so far. The focus has been on a limited set of independent variables (e.g., ontological qualities), dependent variables (e.g., comprehension and understanding), and research methods (e.g., lab experiments). Moreover, CM research has followed assumptions that have been largely stable and unquestioned for decades but are increasingly inconsistent with the IS landscape.

---

[3]For the purposes of this paper, differences in modeling substance versus behavior, form versus change, and static versus dynamic aspects of IS domains are not relevant, as they are all forms of CM. However, differences between these forms exist; they have been discussed elsewhere (e.g., Burton-Jones and Weber 2014; Vessey and Conger 1994).

To substantiate this assertion, we identified the most prevalent *field assumptions* the CM research community has explicitly or implicitly adopted, as evidenced by published research. Field assumptions are broad sets of beliefs about central aspects of a subject that are shared by CM researchers even if they come from different schools of thought (Alvesson and Sandberg 2011, p. 255), such as language/action (e.g., Beynon-Davies 2018; Eriksson et al. 2019), logic (e.g., Clarke et al. 2013), cognition (e.g., Figl et al. 2013), or representation (e.g., Burton-Jones et al. 2017). All schools seem to agree that CM entails building "scripts" with "grammars" that users need to be able to "understand"; they mainly disagree on how to guarantee "good" scripts.

Challenging field assumptions has the potential to generate novel theory and help researchers explore and test new ideas, which is our ambition with this paper. It is also important to review field assumptions to ensure that research remains relevant to practice (Hirschheim 2019). Following Alvesson and Sandberg's (2011) methodology, we conducted a structured, narrative literature review (Paré et al. 2015) of CM journal papers, involving four main steps.

First, we identified a domain of literature (journals that publish CM research). We sampled papers published in the AIS basket of eight journals because these papers are considered to be mainstream, high-quality research in the IS field. To ensure the robustness of our review and to accommodate the view of CM as a niche topic, we also included papers published in the *Journal of Database Management*, *IEEE Transactions on Software Engineering*, and *Information Systems*, which are among the leading substantive publishers of CM studies.

Second, we performed a full-text search of all the papers in these journals between January 2002 and October 2016. We used keywords like "conceptual modeling," "conceptual model*," "conceptual modeling grammar," "ontology," and meaningful variations of these terms. As a temporal boundary to our search, we used the publication date of the CM research framework by Wand and Weber (2002) because it is a path-defining paper for CM[4] that introduced clear definitions of CM concepts such as grammar, script, method, and context.

We examined each paper's title and abstract to exclude papers that were not substantively CM research. Most of the papers we excluded used "conceptual model" to refer to a theory or research framework. This process reduced the total number of CM papers in our sample to 237. We reviewed these papers to confirm their relevance to our study and excluded papers that defined conceptual models as programs or codes (e.g., Krishnan et al. 2004) or used CM concepts to define other aspects of IS, such as the effective use of IS (e.g., Burton-Jones and Grange 2013). This process left us with 197 papers for our analysis (Table 1).

Third, we developed a coding scheme, shown in Appendix A, which evolved in multiple iterations over the course of our research. For example, we developed an "assumptions" dimension once we had decided on the focus of our review.

Fourth, we categorized all papers through an inter-coder process. One author coded all 197 papers while a second author independently coded a random subset of 30 percent (59 papers). The inter-rater reliability scores between the two coders were 91 percent for the raw agreement score and 81 percent for Cohen's Kappa score. The two authors then discussed disagreements, updated the coding criteria and instructions, and independently revised the coding over two more rounds until full agreement was reached. The author who coded all 197 papers then revised the coding of the remaining 138 articles using the updated, agreed coding criteria.

Our review of CM research between 2002 and 2016 is summarized in Appendix B. Our findings demonstrate that the IS community has contributed to the understanding of the practice of CM. In doing so, the community has adopted a largely shared view of the phenomenon and has broadly pursued similar research interests, agreed on key outcome variables, and shared an interest in a set of input factors.

Our interpretation of the community's efforts to date is that the knowledge production in CM research up to this point is coined and bound by four central field assumptions. Table 2 summarizes these assumptions and lists the challenges to these assumptions that are due to digitalization of IS reality (Baskerville et al. 2020; Faulkner and Runde 2019; Tilson et al. 2010; Yoo et al. 2010). With *IS reality*, we refer to the totality of everything that is fundamentally or tangentially related to IT in everyday human experiences (Baskerville et al. 2020). This view spans the traditional activities, resources, and assets that are involved in corporate IT use to include the broader, non-commercial use of IT in society, that is, everything people do to sustain their daily lives that is mediated by digital technologies (Yoo 2010). With a *digital world*, we refer to the totality of the elements that compose the sur-roundings of daily experiences that are enabled by or embodied in digital technologies (Yoo et al. 2012). Finally, Table 2 summarizes the updated assumptions we propose.

---

[4]We conducted informal interviews at the 2017 SIGSAND symposium, where participants largely agreed that Wand and Weber is a seminal publication that guided the development of the CM research community in IS.

**Table 1. Literature Search Results**

| Journals | Initial Search Results | Papers Retained after Screening | Papers Included in Review |
|---|---|---|---|
| *Journal of the Association for Information Systems* | 500 | 18 | 18 |
| *European Journal of Information Systems* | 527 | 19 | 15 |
| *Information Systems Research* | 261 | 18 | 9 |
| *MIS Quarterly* | 659 | 6 | 6 |
| *Information Systems Journal* | 360 | 8 | 5 |
| *Journal of Strategic Information Systems* | 199 | 3 | 2 |
| *Journal of Management Information Systems* | 582 | 6 | 1 |
| *Journal of Information Technology* | 265 | 2 | 1 |
| *IEEE Transactions on Software Engineering* | 416 | 58 | 54 |
| *Journal of Database Management* | 326 | 53 | 49 |
| *Information Systems* | 127 | 46 | 37 |
| **Total** | **4,222** | **237** | **197** |

**Table 2. Conceptual Modeling Assumptions**

| Type of Assumption | Traditional Assumption | Challenge to Assumption | New Assumption |
|---|---|---|---|
| *The representation assumption* | Scripts represent physical reality. | Human experience is increasingly, at least partially, computed.<br><br>IS increasingly not only represents but also creates, shapes, and governs physical reality. | Scripts represent both physical and digital realities, and mediate transitions between the two. |
| *The structure assumption* | Scripts represent the deep structure of IS. | Human experience increasingly involves digital objects, which blur the distinctions between physical, deep, and surface structures of IS. | Scripts represent deep structure and its couplings with physical and/or surface structure of digital objects. |
| *The agency assumption* | Scripts are produced and consumed by humans. | Digital objects increasingly have material agency. | Scripts are produced and consumed by both human and digital agents. |
| *The context assumption* | CM is a professional activity that occurs in organizational work settings. | IS are increasingly developed and deployed not only in organizational work but also in the nonwork settings of everyday life. | CM occurs inside and outside of organizational work settings. |

## The Representation Assumption

**Old assumption: Scripts represent physical reality**. One key field assumption of CM research is that CM is for developing scripts to represent physical reality. Our review (Table B3) revealed that 95 percent of CM papers focused on representing properties, states, structures of, and interactions between material and socially constructed things in the physical world alone. This finding was expected since the common definitions of CM refer to creating graphical scripts to represent a real-world domain that an IS supports (Recker et al. 2019; Wand and Weber 2002).

Because of this focus, CM research has often built on works of ontology, that is, the study of the nature of the world and attempts to organize and describe what exists in reality (Burton-Jones and Weber 2014; Green and Rosemann 2004).

With this focus, evaluations of CM scripts have focused on measures like representational accuracy (Kimelman et al. 2009; Kounev 2006) and representational quality (Currim and Ram 2012; Koschmider et al. 2010; Recker et al. 2010). These measures have been employed to evaluate how "completely and clearly" a script represents relevant phenomena in the real world (e.g., Shanks et al. 2010; Shanks et al. 2008). The use of ontology in CM research has been so prominent that it became a reference for research in other domains, such as data quality (Price and Shanks 2005; Wand and Wang 1996).

While works on ontology have been dominant, other theories have also been used to provide guidelines to facilitate the creation and improvement of CM scripts. However, most of these theories also assume that CM scripts describe the parts of the physical reality that an IS must represent. For example, Figl et al. (2013) and Mendling et al. (2010) used cognitive theories to evaluate the transparency of a CM script in terms of how closely the script's visual symbols matched referent real-world concepts.

**Challenge to assumption: Digital objects increasingly shape reality**. Digital objects are human-made artifacts that are at least partially computed (Faulkner and Runde 2019). Their structure and behavior can change through (re-) programming of their digital layer (Yoo et al. 2010).

Today, digital objects are part of many first-world human experiences (Yoo 2010). For example, we plan holidays using recommender systems, we choose modes of transportation using shared mobility service platforms, and we engage in biophysical activities in response to signals from wearable devices. These examples show that IS no longer merely maps physical reality representations onto digital objects but increasingly assume a more active role by creating, shaping, and governing physical reality (Baskerville et al. 2020). For instance, procedural generation tools autonomously create landscapes in video games (Seidel et al. 2018), solve design challenges in the manufacture of semiconductors (Brown and Linden 2011), and produce high-quality visual content (Huang et al. 2016).

Digital objects do more than just create new objects. They continually shape both digital and physical realities. For example, self-driving cars operate within a physical domain and must recognize and interact with certain physical objects, be they pedestrians, regular cars, or obstacles. At the same time, how a self-driving car constructs and interprets its own digital reality can affect its adherence to legal, ethical and moral norms and how it learns traditional and emergent traffic behaviors. The physical reality in which the vehicle operates changes through how it interprets its digital reality, just as much as changes in the physical reality, such as new traffic patterns, new opportunities for commercial and public transportation, and new rules and regulations, lead to changes in its digital reality.

**New assumption: Scripts represent both physical and digital realities and mediate transitions between the two**. CM's traditional purview of creating scripts of physical reality must be broadened so CM is also about creating scripts of digital reality in its own right. This shift in scope will help to explain the ever-expanding and increasingly complex digital world and facilitate change in either the physical or digital world.

Moreover, rather than seeing CM as mainly guiding the creation of *representations* of physical reality in digital artifacts (e.g., a database), we contend that a key function of CM in an increasingly digital world is *mediation*. By mediation we mean a structured process of facilitating understanding, communication, and change to capture, share, and translate among aspects of multiple realities.[5] In this understanding, CM provides tools that humans and digital agents can use to *mediate the transition between various states of digital and physical realities*. The key insight here is that CM assists in or enables these transitions. Facts, assumptions, and beliefs gleaned through CM can be used to understand relevant elements of reality (physical or digital) *and* institute change in either the digital or the physical world.

### *The Structure Assumption*

**Old assumption: Scripts represent the deep structural meaning of IS.** Our literature review (Table B3) shows that 89 percent of CM papers have focused on representing the deep structural meaning of IS, which Weber (1997) suggested lies at the core of the IS field. Therefore, as Weber contended, the question of the quality of a deep structure representation should be a major focus. However, over time a dominant view emerged that CM should be used to represent *only* deep structure phenomena, even though Weber pointed out that both surface and physical structure phenomena should also be featured in CM to describe the choices that designers made. In our review, only two papers were concerned with using CM to represent the surface structure aspects of IS, one of which examined how OLAP schema interfaces could be personalized according to users' needs (Garrigós et al. 2012). Similarly, we found only three papers that used CM to repre-sent the physical structure aspects of IS. For example, Vergara et al. (2007) examined how model-driven web engineering can accommodate the interoperability constraints that are imposed through external, distributed assets that rely on proprietary technical infrastructure standards like adaptors or protocols.

---

[5]Our definition of mediation is inspired by Latour's (2005, p. 39) notion of a mediator as an entity or actor with the ability to "transform, translate, distort, and modify the meaning or the elements they are supposed to carry."

**Challenge to assumption: Digital objects blur the distinctions between deep, surface, and physical structures**. Digital objects are enduring, structured entities (Faulkner and Runde 2019) that are composed of data and metadata and regulated by structures or schemas (Hui 2016). Because they are layered and modular (Yoo 2010), digital objects can be software but can also be embedded in material objects such as toothbrushes, smartphones, cars, and airplanes.

As a consequence and in contrast to the general-purpose hardware that underpinned most traditional IS (e.g., PCs and servers), the characteristics, constraints, and flexibility of new types of hardware and software are central to how many digital objects are used. Many digital objects involve specialized hardware that can connect seamlessly with real-world behaviors, and software that can expose the data that specialized hardware captures in a way that is meaningful to the user, depending on the interface. In analyzing or designing such digital objects, developers must consider hardware requirements to foster the deep meaning of user interactions. At the same time, software must be developed such that it responds to real-world interactions and captures, translates/interprets, and displays meaning to users.

To illustrate, consider a wearable device like Fitbit. The device's pedometer, a physical structure element, directly captures one kind of data (the deep structure element) the device was designed to manage while at the same time constraining the shape and form of these data to the capacities of the pedometer. The absence of a large screen (a surface structure element) impacts how users can interact with the device, which is primarily through finger tapping. Both the surface and the physical structures shape Fitbit's deep structure. For example, its semantic schema must be able to process different tapping rhythms and frequencies.

**New assumption: Scripts represent deep structure and its couplings with the physical and/or surface structure of digital objects**. CM's traditional focus on capturing the deep structure of IS must be broadened to include the physical and/or surface structure of digital objects, as well as the couplings between the structures. The meaning of digital objects is no longer solely vested in their deep structure but emerges from how material and non-material components are combined into hybrid forms (Faulkner and Runde 2019).

With this updated assumption, CM must consider any relevant structural properties of objects in physical or digital reality. For example, CM may be used to represent not only the structure of a domain but also facts about the physical medium in which a digital object will reside. Likewise, CM may be used to represent multimedia and graphic elements of a user interface.

## The Agency Assumption

**Old assumption: Scripts are produced and consumed by humans**. A third field assumption we found is that CM is a task undertaken and directed by *humans* as producers and consumers of CM scripts. Of the 197 papers we reviewed, 176 examined CM as a task undertaken by humans in such roles as professional IS analysts (172), end users (2), or students in learning and teaching scenarios (2) (Table B3). Only 18 papers (10%) focused on nonhuman, digital agents involved in producing or consuming scripts. In these papers, the focus was on developing new algorithms for CM tasks like script creation (e.g., Purao et al. 2003), script validation (e.g., Eshuis and Wieringa 2004), or script transformation (e.g., Burgueno et al. 2014).

The dominant view of CM as a human activity is deeply engrained in the CM literature. For example, Topi and Ramesh (2002) argued that CM research focuses mostly on humans' developing and using CM scripts to communicate between analysts and developers. In line with this focus, most CM studies that have examined human participants characterize them through social attributes like their level of training, technical experience, task-related experience, domain experience, or modeling experience (Recker 2010; Reijers et al. 2011). Many CM papers evaluate how individual differences in human expertise affect CM practices (Mendling et al. 2019). We found no corresponding characterizations of digital agents (e.g., algorithms) that are involved in CM activities.

**Challenge to assumption: Digital objects increasingly have material agency**. Digital objects increasingly have material agency, that is, the capacity to act on their own, without human intervention (Leonardi 2011). Robots, autonomous vehicles, facial-recognition software, natural-language processing tools, virtual bots, and machine learning platforms increasingly perform versions of the kind of cognitive functions that are typically associated with humans, such as perceiving, reasoning, learning, and interacting (Rai et al. 2019). Such digital objects do not just inform or automate (Zuboff 1985). They start to assume mastery and control. Examples include software-controlled vehicles that drive autonomously (Frazzoli et al. 2002), roboAdvisor software that manages investments by automatically rebalancing portfolios based on target allocations (Lee and Shin 2018), and AI underwriters that process loans (Markus 2017).

With these capabilities, digital objects increasingly feature as agents in their own right. For example, Apple's Siri, IBM's Watson, and Google's Waymo can be seen as autonomous entities. Even though they implement certain algorithms in a predictable manner and their behavior is determined by written code, they routinely make autonomous or semi-autonomous decisions, and they do so in increasing numbers of domains. How these digital entities "understand" the world and increase their ability to act in it must therefore become an important concern for CM because cognition rests on representation (Edelman 1999; Sowa 1999).

**New assumption: Scripts are produced and consumed by both human and digital agents**. The conception of agency in CM must be relaxed. So far we have assumed that scripts are produced and/or consumed by humans who have proficiency and expertise in CM methods and domains. But scripts can also be produced and/or consumed by digital agents, meaning nonhuman, at least partially computed, artifacts like algorithms, autonomous tools, bots, APIs, and generative engines. Digital objects already outnumber humans as information processors. In 2016, more than 20 billion devices were connected and leveraged more than 50 billion sensors (Zhang 2016). CM scripts can already be constructed through machine learning algorithms that mine digital trace data (van der Aalst 2016) or through natural-language processing algorithms that can interpret verbally expressed requirements (Friedrich et al. 2011). Likewise, algorithms and execution engines already interpret and execute CM scripts (Ouyang et al. 2009) and perform mapping between scripts (Malavolta et al. 2009) so the scripts can be consumed by other digital agents, such as APIs or execution engines. As advances in artificial intelligence continue, the variety and level of digital objects' agency are set to increase as more digital agents produce scripts of physical and digital realities, and more digital agents interpret and execute them.

### The Context Assumption

**Old assumption: CM is a professional activity that occurs in organizational work settings**. The last major field assumption that our review made evident is that CM is a profession-related exercise that is carried out within the boundaries of organizational work. More than 97 percent of the CM papers we reviewed investigated how CM scripts can be developed or used within organizational work contexts but not beyond (Table B3). The reported purposes of CM include representing organizations' business processes (Recker et al. 2009), increasing the awareness of and knowledge about business operations (Bandara et al. 2005; Recker et al. 2010; Samuel et al. 2015), deconstructing organizational complexity and managing organizational change (de Albuquerque and Christ 2015; Weidlich and Mendling 2012), and specifying data's requirements, structure, and use in organizations (Ågerfalk and Eriksson 2004; Bowen et al. 2009). In total, 87 percent of the papers in our review portrayed CM as a professional exercise (e.g., Bera et al. 2014) that is carried out by an organizational work force that is trained in CM, business, or both (e.g., Burton-Jones and Meso 2008; Milton et al. 2012). Typical work roles mentioned in CM papers include consultant, analyst, or designer (e.g., Milton et al. 2012; Recker et al. 2011).

There are good reasons for this assumption. The IS discipline emanated from business and management schools (Hirschheim and Klein 2012), leading to research primarily having an organizational focus (Yoo 2010, p. 215). Likewise, CM has traditionally been developed during the early phases of systems analysis and design to elicit, define, analyze, and communicate system requirements between members of the IT profession and other members of the organizational workforce (Burton-Jones and Meso 2006). Empirical surveys have indicated that more than 70 percent of users work with CM scripts to clarify domain and IS requirements among the members of the organizational design team (Dobing and Parsons 2008).

In sum, the CM studies to date have assumed that IS are analyzed or designed to operate in organizational work contexts (e.g., Krogstie et al. 2006; Weidlich and Mendling 2012), so CM scripts are viewed as integrative communication tools (Parsons 2002) that allow members of organizations to relate their views of a domain to other views in the organization.

**Challenge to assumption: IS are increasingly developed and deployed not only in organizational work but also in nonwork settings of everyday life**. Digital objects now feature in many everyday activities that we do to sustain our daily lives, beyond organizational work settings alone (Yoo 2010). Not only do we use digital technologies like computers and laptops at work, also we use these and other technologies (such as smart, connected devices, online social networks, and wearable devices) long after we have left work.

This increase in IS deployment is matched by a widening scope of IS development. IS are being designed and built but not only for organizational work but also for purposes outside it, such as user-generated social media content, personal website design, and end-user data management.

IS development as part of everyday life beyond organizational work alone means that IS development is no longer the sole purview of IT departments. Almost everyone can develop IS for their personal use. Development tools like Wordpress.com, Airtable, and Webflow.com expand the reach of IS development to nonexpert cohorts, including children (Kraleva et al. 2019), through new practices like drag-and-drop programming. Working with these tools, even those with little to no specialized knowledge can build their own websites, apps, productivity software, and data-management solutions. They implicitly or explicitly engage in CM when they learn to program robots using Lego Mindstorm, build a personal website structure using Wordpress, or design responsive database structures using Webflow.com. Undergirding these trends are changes in primary school curriculums, which have begun to emphasize a variety of computer literacy skills, including programming and software development (Saçkes et al. 2011). Open-source software development (Bagozzi and Dholakia 2006; Stewart and Gosain 2006), where users outside the traditional organizational workforce develop software and use CM alongside IT professionals, provides another illustration of these trends. These initiatives involve CM to a much larger extent than typically assumed. Robles et al. (2017) identified over 93,000 UML scripts across 24,000 projects on GitHub, a prominent open-source software development platform.

**New assumption: CM occurs inside and outside of organizational work settings**. CM is no longer only an organization-focused, professional work activity. Instead, the production and consumption of scripts takes place both inside and outside traditional organizational boundaries, and involves both professional and nonprofessional agents. Tools, platforms, and systems that enable or support CM production and consumption activities are designed and used in a wider social context than the organizational, professional work setting alone.

With this wider reach, CM users include organizational staff like trained IS professionals, as well as ordinary individuals outside of the organizational context (Yoo 2010, p. 217). This wide heterogeneity of users increases both the diversity of objects to be represented and the differences in CM usage patterns (Lukyanenko et al. 2017). For example, in some settings, users are no longer passive content consumers but are also content producers (Ritzer et al. 2012). Collective action movements in government, science, and business involves the public in decision-making, data collection, and analysis. Known by such labels as "citizen science" (Levy and Germonprez 2017), "crowdsourcing" (Bayus 2013), and "participatory governance" (Pellizzoni 2003), such collective action is fueled by the promise of novel, out-of-the-box perspectives that ordinary individuals can provide. As Lukyanenko et al. (2016, p. 447) observed, "because citizens generally lack formal … training, they view problems and issues in light of their own knowledge and interests, creating fertile ground for discoveries."

### Interim Conclusions from the Literature Review

The CM community has generally worked within its shared field assumptions, resulting in a cumulative research tradition with a shared set of methods and instruments that has produced a constant stream of high-quality papers. Most studies have built on theories of ontology (e.g., Burton-Jones and Meso 2006; Milton et al. 2012) or cognition (e.g., Evermann 2005; Figl et al. 2013) to define the antecedents of outcome variables, such as script comprehensibility (e.g., Parsons 2011; Recker 2013), domain understanding (e.g., Bera et al. 2014; Burton-Jones and Meso 2006), and similar measures of representational quality (e.g., Genero et al. 2011; Poels et al. 2011). The studies have focused on organizational settings involving professional workers as producers or consumers of CM scripts.

The embeddedness of assumptions in CM research is not a limitation *per se*. Field assumptions provide valuable structuration mechanisms that allow researchers to build incrementally on others' work, supporting the accumulation of knowledge. They also accelerate research processes because they provide shared conceptual lexica, measures, instruments, materials, and procedures.

However, assumptions can become detrimental (Feyerabend 2010; Kuhn 1991) if they create stereotypes that increase research within the bounds of existing assumptions at the expense of research outside those assumptions (Gray and Cooper 2010). They can constrain the breadth of CM research by limiting opportunities to explore unchartered territories, generate new theories and hypotheses, and increase the empirical scope.

## A New Framework for Conceptual Modeling ▬▬▬▬▬▬

Our review shows that the CM field assumptions have been largely stable, even though the IS landscape has not. Developments in technology and organizing have made CM field assumptions increasingly distant from and poorly fitting with IS reality. These developments limit the contributions and explanatory power of CM research that remains within the realm constructed by the old assumptions. If research communities cling to and propagate outdated views, they lose relevance and credibility (Hirschheim 2019; Hovorka et al. 2019). Therefore, a change of direction in CM research is timely and warranted. In response, we propose a novel framework for CM research and update its research agenda. The new perspective we propose is shown in Figure 1 as the conceptual modeling in the digital world (CMDW) framework. The key constructs that conceptualize the elements of the framework are described in Table 3. The updated definitions of the constructs mirror the update in assumptions shown in Table 2.

In constructing the framework, we remained cognizant of almost five decades of research on CM and accounted for the conceptual foundations that underlie CM research and practice. Thus, at its core, the CMDW Framework presents the four key CM constructs distilled by Wand and Weber (2002): the CM script, method, grammar and context. For example, in the new framework, a CM method still describes the procedures by which a CM grammar can be used, and which could be implemented as rules and principles and supported with tools and techniques. A CM grammar in the new framework is still a set of constructs and rules that show how to combine these constructs, and it still has syntactic and semantic elements (Burton-Jones et al. 2009). To maintain the conceptual lineage, Figure 1 visualizes the constructs in the CMDW framework graphically, analogous to Wand and Weber, but the roles and relationships between the constructs are conceptualized differently in several ways.

The first difference in our conceptualization is that our framework recognizes that CM occurs within and between physical and digital realities, as indicated visually in Figure 1 through two separate but partially overlapping shaded areas within the CM context (Baskerville et al. 2020).

The second difference is that our framework situates the *CM script* as the principal CM artifact at the intersection of physical and digital realities because it serves multiple functions in mediating between these realities. These functions are indicated visually in Figure 1 through four different, labeled arrows. This conceptualization stresses that mediation involves representation (of things in the physical or digital reality), as well as translation (e.g., from digital to physical reality), execution (e.g., within digital reality), and change (e.g., states of things in physical reality).

The third difference is that we visually place the script construct *outside* the method and grammar constructs in Figure 1, while Wand and Weber (2002) conceptualized the CM script as an outcome bounded by a CM grammar and method. This conceptualization provides for the possibility of decoupling scripts from CM methods and CM grammars. For example, nonexpert developers might create nascent representations that do not adhere strictly to the rules of existing grammars.

The fourth difference is that our framework extends Wand and Weber (2002) by recognizing the *CM agent* as any human or digital agent, not just human agents, who/that produces or consumes CM scripts (highlighted visually through placement at the overlap between physical and digital realities).

Finally, our framework offers a new definition of the *CM context* by describing the CM context broadly as the continually changing socio-material setting in which CM occurs (Table 3). This setting is distributed over time and space and involves physical as well as digital objects and events across digital and physical realities. It occurs both within and between organizational work settings and the nonwork settings of everyday life. This expanded definition updates Wand and Weber's (2002) definition by emphasizing that CM activities are no longer performed within organizational, professional contexts alone but can be undertaken in any organizational or non-organizational setting. This broader context, we argue, provides CM scholarship the opportunity to go from niche to mainstream.

## Discussion

We now discuss the key shifts in perspectives on CM propagated by the CMDW framework. They flow from our reconceptualization of existing constructs (e.g., script or context) and our addition of new CM constructs (e.g., agent, digital reality, or mediating relationships).

### *The CM Script Is the Focal CM Artifact*

In CM research to date, considerable emphasis has been on grammars and methods (Table B1). In contrast, we position the CM script as the core *CM artifact*. Our argument is that the CMDW framework does not insist on a CM script being an instance of a particular grammar.

With end-user computing increasing and CM opening to a broader population, a strict coupling of scripts to grammars is no longer realistic. In a realm in which millions of people may be engaged in IS development and deployment, new forms of CM may emerge that may not comply with any existing grammars. Nonexpert communities already model domains of physical reality using representational methods that do not conform to any established modeling grammar. For example, to communicate the intricacies of the possible ties between the U.S. President, Donald Trump, and his Russian counterpart, Vladimir Putin, journalists have been using a variety of CM scripts that draw on the grammatical elements of entity-relationship, process, network, and semantic models, infused with rich multimedia elements (Yourish and Buchanan 2019). IS practitioners recognize the versatility of such *ad hoc*, hybrid CM scripts in some cases and even recommend using them over traditional, more formalized modeling grammars like ER or UML (Frisendal 2016). Studying the development and qualities of such CM scripts in their own right may uncover innovative solutions that precipitate the development of formalized CM grammars of the future.

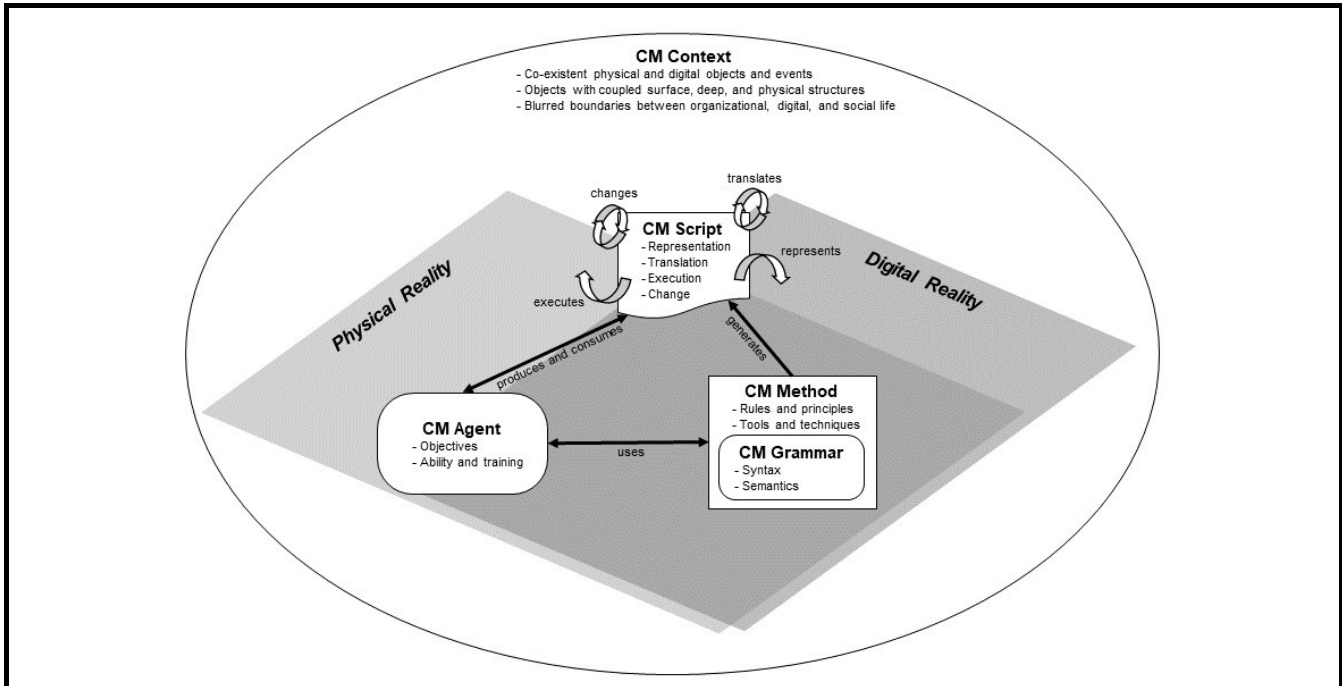

**Figure 1.  A New Framework for Conceptual Modeling in the Digital World (CMDW Framework)**

| Construct | Wand and Weber's(2002) Definition | Updated CMDW Definition |
|---|---|---|
| **Table 3.  Construct Definitions in the CMDW Framework** | | |
| *CM Script* | A statement generated in the language of a CM grammar that provides a description of the real-world phenomena that an IS is intended to represent. | A generated statement that is suitable for purposes of mediation and provides a description of the phenomena of a physical and/or digital reality. |
| *CM Method* | The procedures by which a CM grammar can be used. | *Same* |
| *CM Grammar* | A set of constructs and rules that shows how to combine the constructs to model real-world domains. | A set of constructs and rules that shows how to combine the constructs to model physical or digital domains of reality. |
| *CM Agent* | Not explicitly defined.  Implicitly assumed to be a human agent that produces and/or consumes CM scripts | A human or digital agent that produces and/or consumes CM scripts. |
| *CM Context* | The setting in which CM occurs and scripts are used, including individual difference, task, and social agenda factors. | The intertwined physical and digital reality setting in which CM occurs and scripts are produced and consumed. |
| *Physical Reality* | The aggregation of constituent material and socially constructed things and their properties that exist in the real world. | *Same* |
| *Digital Reality* | | The aggregation of logical and non-material things and their properties that exist in the computed, digital realm. |
| *Mediation* | | Activities related to facilitating representation, translation, execution, and change between aspects of physical and digital realities. |

This separation of CM scripts from formalized CM grammars is consistent with modern linguistic theory. Since all CM scripts are linguistic expressions, they can be *grammatical* (Chomsky 2002), that is, valid based on the rules and constructs of a particular grammar, such as English or UML. For example, two named boxes connected with a line are grammatical for numerous CM grammars, including various notations of ERD and UML. However, CM scripts can also be *ungrammatical*, that is, in violation of a given grammar. For example, the objectification of unary facts is ungrammatical for ORM-1 (Halpin 2006). Producing a script that violates a CM grammar is often done in teaching (such as in student exercises that ask to find grammatical errors in invalid scripts, see as an example Halpin and Morgan 2008, p. 94). It also occurs in practice, as when ungrammatical guidelines are introduced to override grammatical rules (Recker et al. 2010; Samuel et al. 2015).

Ungrammatical CM scripts do not come from nowhere; they are products of some generative mechanisms. In linguistics, all expressions are assumed to be instances of some underlying generative process (some basic or universal grammar), so studying the properties of this underlying generative system has been a major focus of modern linguistics (e.g., Chierchia and McConnell-Ginet 2000; Lyons 1991). Our framework recognizes that such mechanisms could also have properties of theoretical interest to CM scholars by suggesting that CM research should be interested in both grammatical CM scripts (the traditional CM perspective) and ungrammatical CM scripts (a novel idea, to our knowledge). We promote the use of valid, grammatical CM scripts. Grammatically valid scripts ensure consistent communication of the messages and a reliable execution of its mediating functions. At the same time, script violations should also become valuable objects of CM research. Ungrammatical CM scripts may carry important signals regarding how CM is actually used and could reveal some of the basic characteristics that underlie all CM grammars.

## CM Scripts Perform Mediation Functions Beyond Representation Alone

A second key perspective propagated by our framework is that the role of CM extends beyond representation to include mediation. As physical and digital realities become increasingly entwined, and the digital world grows ever more complex, a major challenge will be to facilitate the transition between states of reality. Our position is that CM scripts are well-positioned for this task.

Positioning CM scripts as mediating objects is fully consistent with the traditional view of CM scripts as representational artifacts. Representation (e.g., of the states a material thing traverses in physical reality through a CM script) is a key element of mediation. However, mediation implies activities beyond mere representation, as CM scripts must be situated in the realities (physical and/or digital) to support the transition of a modeled phenomenon between states of reality. Figure 1 shows CM scripts have four functions to perform, instead of one:

1. **Represent physical reality in digital reality**. CM scripts have always been tools that translated assumptions about physical reality into a form and content suitable for the development of digital software components, such as databases or program code (Mylopoulos 1992). CM remains central to ensuring that physical reality is appropriately represented in the features and processes of IS embodied in or enabled by digital objects.

2. **Execute digital reality within physical reality**. Digital objects and events increasingly shape physical reality, signaling presence of ontological reversal (Baskerville et al. 2020). A good example is the use of OpenSCAD scripts during 3D printing (Kyriakou et al. 2017). These scripts capture digital designs that can be transformed into physical objects by representing the digital object in a way that human users can understand and reuse the design (Kyriakou et al. 2017). They must also carry precise semantics that can instruct the digital printing machinery (e.g., components like extrusion nozzles, thermal bedding, and filament detectors) about how the design should be sliced and how each layer the slicer generates is to be printed.

3. **Translate between digital realities**. As the digital world grows, the number of digital CM agents that produce and/or consume CM scripts will increase. Examples include application programming interfaces (APIs), remote procedure calls, smart contracts, and bots. When digital agents interact with each other, CM scripts translate between the languages that specify how each digital agent operates. For example, consider popular AI tools labeled "deepfakes," which create convincing, yet entirely fake, pictures of people using algorithms that generate image representations from available textual descriptions (Porter 2019). These digital agents use so-called adversarial CM scripts that compete against each other in a MinMax two-player game (Zhang et al. 2019). The scripts specify representations of a textual real-world object in such a way that a second script fails to distinguish between the digital representation and the physical referent object. The agents produce multiple script translations (from text to picture and from picture to text) that compete with each other.

4. **Change physical reality**. In the past, CM scripts have dealt primarily with the deep structure meaning of an IS. With the blurring of deep, physical, and surface boundaries, CM scripts must now permeate more aspects of the IT infrastructure, including physical materials, code, and interfaces. This situation creates avenues for CM scripts to assist with changes between states of physical reality. Consider digital objects like wearable devices that are designed to change physical behaviors (Bonfiglio and De Rossi 2011). Many of these devices are designed to assist user groups (e.g., chronic patients) in adapting their behaviors (e.g., eating healthier). Representing and processing physical data (e.g., live biofeedback from neurophysiological sensors) in real time is, therefore, an important design challenge (Lux et al. 2018), while behavioral change in actual human experience is the most important outcome (Van Woensel et al. 2015). Therefore, how physical data, such as steps, calorie intake, or heart rate, is modeled in a script is an important aspect of facilitating change in physical reality, such as attaining healthier shopping habits or consuming more nutritious food (Kim 2014).

We propose that CM scripts are well positioned to embrace this wider array of mediation functions. CM scripts have always been designed to span boundaries between people and realities. For example, to verify requirements, technical developers routinely shared CM scripts with business users to bridge gaps in technical expertise, knowledge, and language (Bera et al. 2014; Dietz and Juhrisch 2012). CM scripts were also used to reconcile the differing values and beliefs of users, departments, or even organizations (Parsons 2002; Soffer and Wand 2007).

## CM Scripts Are Produced and Consumed by Social and Digital CM Agents

A third key perspective our framework takes is that of the CM agents who/that produce and/or consume CM scripts. Research has largely assumed that CM agents are humans. CM research has also restricted their role to that of a control variable (Mendling et al. 2019). Our new framework situates agents as a focal object of interest and classifies these agents as human or digital, and professional (e.g., experienced analysts or trained algorithms) or unprofessional (e.g., ordinary individuals or untrained algorithms).

These distinctions are important. Digital agents as producers or consumers of CM scripts have already been the focus of some CM research, but no conception of common underlying traits or differentiating characteristics has yet emerged. For example, digital agents that operate on the basis of supervised machine learning algorithms will produce CM scripts whose quality differs from that based on unsupervised machine learning algorithms.

The key objective for digital agents in consuming CM scripts also differs from that of human agents. Humans seek to glean complete and clear understanding of a domain from CM scripts. Digital agents, such as workflow execution languages, have no such objective. Their documented objectives at this point include script construction (Friedrich et al. 2011), validation (Montero et al. 2007), transformation (Malavolta et al. 2009), and analysis (Leopold et al. 2014). As digital technologies become more intelligent, additional objectives may emerge.

The emphasis on the CM agent in the CMDW framework also extends to human agents. Emphasis in extant CM research has largely been on professionally trained organizational workers. The CMDW framework also includes human agents with differing levels of cognitive ability and training. We need to better understand ordinary and untrained human script designers, including the challenges they face when they create CM scripts, the assumptions, beliefs, and design knowledge they bring to bear, and the kinds of CM scripts they *already* use to accomplish their tasks. To illustrate, consider *citizen modeling* (Lukyanenko et al. 2017), that is, the question concerning how the vast and diverse views and data content that citizen science contributors generate can be captured in an IS. In the prominent citizen science project Fold.It (Koepnick et al. 2019), any interested member of the public can use a virtual simulator to fold protein structures. Individuals can create any conceptual structure they desire, including realistic ones, those that match existing structures in the physical world, or other structures that are impossible to conceive based on current knowledge and technologies. Scientists then analyze selected scripts to determine whether these new structures can be applied to eradicate diseases (including COVID-19[6]) or to create other biochemical innovations.

---

[6]https://fold.it/portal/node/2008931

## CM Occurs in a Broader CM Context Situated in Intertwined Physical and Digital Realities

Our framework distinguishes between digital and physical objects and events in physical and digital realities. As Figure 1 shows, we see this distinction as blurred. For example, digital objects rest on physical objects (Faulkner and Runde 2019) because digital bits are stored in some physical medium and are processed using electric signals. Likewise, physical reality is increasingly punctured by objects and events from digital reality. The intertwining of digital and physical realities will only increase as humans rely more on extended cognition, that is, physical, social, and digital cognitive systems outside of the individual organism (Clark and Chalmers 1998).

Digital reality will also be increasingly prominent as a contextual setting in which CM occurs. While grounded in the physical world, digital reality is increasingly a world of its own and has unique properties. For example, in video games digital agents operate on the basis of scripts to generate a reality that humans playing the game will experience (Seidel et al. 2018).

The distinction between physical and digital reality in our framework emphasizes the growing importance of the digital world to our physical reality and the unique identity of CM scripts as tools for transitioning between the worlds. To illustrate, consider mediations between digital and physical reality that occur in what is known as "augmented reality." An example is Pokemon Go, a mobile application through which players try to capture digital objects in their physical environments. Powered by augmented reality technology that fuses a digital world with the physical world, Pokemon Go allows players to traverse the physical world following a digital map to respond to events (e.g., spawning of Pokemon creatures) that occur in the digital world. The game can also overlay digital imagery on a person's view of the physical world using a smartphone screen, making it appear as though, for example, a turtle is swimming across one's living room. The way the digital world is modeled in Pokemon Go's script thus both informs and shapes behavior in the physical world. Engaging with Pokemon Go reportedly results in elevated physical activity levels (Althoff et al. 2016).

Calling attention to a broader view of a CM context is congruent with the growing interest in theorizing about context in IS (Avgerou 2019; Hong et al. 2014). CM research should be more sensitive to the context in which CM occurs. Digital objects (which may be created with the help of CM) can alter the perceptions, beliefs, and behaviors of CM agents in their physical or digital realities, which may, in turn, influence any subsequent CM activity in which the agents may engage. To begin with context-sensitive research, we encourage research on the digital CM context, such as virtual realities used in massive multiplayer games or market simulations, because it is an uncharted territory for CM scholarship. These new contexts promise to reveal new applications and use cases for CM and will likely feature new forms of CM scripts.

# How CM Research Should Continue ■

The main value proposition of our framework is that it repositions CM scholarship in the digital world so it can continue its prominent role in IS scholarship and heighten its significance in an increasingly digital world.

A second core value proposition of our CMDW framework lies in highlighting uncharted prospects of CM research. These prospects can reignite interest in CM research. Our framework promotes a new, proactive agenda that encourages CM scholarship to speak to the broader issues and challenges organizations and society face in a digital world. At the same time, our framework stays connected to traditional CM scholarship, easing the transition between traditional CM research and uncharted territory.

To make these contribution claims as tangible as possible, we now discuss three ways in which our framework can guide future CM scholarship. These three ways focus on three fundamental ideas of our framework:

1. In an increasingly digital world, representation of physical reality remains important.

2. One essential function of CM scripts is to translate between digital agents.

3. CM scripts play a central role in facilitating change in physical reality.

## Representation in Algorithmic Analytics

Ensuring that physical reality is appropriately represented in digital objects' features and processes remains a valuable stream of CM research. To illustrate a novel research opportunity in this traditional area of CM scholarship, consider algorithmic analytics for managerial decision-making (Khatri and Samuel 2019). The practice of analytics relies on analyzing large volumes of heterogeneous data about aspects of physical reality (e.g., store operations, sales, shipment of goods) through digital objects (e.g., electronic dashboards, analytics software, data discovery platforms). Beyond the traditional research question of how analytics software could be developed on the basis of a CM script of the business domain, the CMDW framework suggests at least two novel foci for future research, one concerning the CM agent, and one concerning the CM scripts themselves.

In focusing on the CM agent, CM research could examine the beneficial and inhibiting aspects of existing CM scripts for two types of agents involved in making decisions using analytics, the human agent and the digital agent. In terms of digital agents, the question concerns how CM scripts for analytics could be made machine-readable by digital agents such as automated decision-making algorithms. In current practice involving human agents, both CM scripts such as information catalogs and CM methods, such as data lineage and metadata management, are used to guide data engineers and data architects in designing and integrating data for analytics. Ideally, these same CM scripts remain useful for other CM agents (e.g., a data scientist who performs analyses to uncover previously unknown insights) to consume during analytics so they can understand the physical reality that is documented digitally. Dependent variables of interest in research on such CM agents could include not only the understandability of the script or its accuracy, but also the subjective quality of decisions made, the factual accuracy of decisions made (e.g., forecasting), and the novelty of business ideas created.

As for focusing on the CM scripts themselves, algorithmic analytics requires highly formalized, semantically precise scripts for the machine learning algorithms to function. Semantically precise scripts consumed by digital CM agents are not always interpretable to humans, which can undermine their trust in the decisions an algorithm produces (Watson and Nations 2019). In practice, multiple representations, including combinations of text and graphic visualization, are employed to communicate key conclusions from data analysis to a human audience (Elias et al. 2013).

Our framework proposes the need to understand which combination of grammatical and ungrammatical scripts optimizes *both* the effectiveness of decision-making and trust in algorithmic decisions. We also need to understand how formal and informal CM grammars can be used together to represent and communicate key conclusions based on analytic data. For example, a traditional grammatical CM representation (e.g., an ER diagram) might be effective in representing data as input for analytics processes and clarifying assumptions about what data was used and which new analytics insights were made. However, a combination of both text and graphic representations, such as narratives, pictures, and videos, might be better suited to communicating outcomes. Clarifying how combinations of grammatical and ungrammatical scripts facilitate understanding of the results of analytics is a non-trivial CM research objective. Techniques such as local interpretable model–agnostic explanations (Ribeiro et al. 2016) illustrate the continued relevance to practice of this seemingly traditional question of representation.

## Translating Between Digital Agents in Smart Contracts

The second key idea of our framework is that digital CM agents build on CM scripts to operate in digital reality alone, bypassing physical reality. In such settings, digital agents are both consumers and producers of CM scripts.

Smart contracts are examples of digital objects that interact with one another (Christidis and Devetsikiotis 2016). They execute and autonomously enforce digital agreements through CM scripts that include digital representations of real-world objects and events. For example, the script of a smart contract might stipulate that a certain price level of stock triggers a subsequent trade, or that the result of a sporting event executes a bet that was previously defined in a smart contract, which then pays the agreed amount to the bet winner.

As these examples show, smart contracts are digital agents that operate through their scripts, bypassing physical or social involvement. Once the conditions of these contracts have been set in a script, the transactions are executed without the need for human intervention (Cieplak and Leefatt 2017). A CM script called "oracle" serves as a mid-layer technology that provides reliable data for the correct execution of a contract. For example, this CM script could pull the schedules and results of sporting events from a reputable source through an API, look at the conditions of the smart contract, and execute it (e.g., send the agreed prize to the winning party through an online transaction service).

Our framework suggests that effective use of smart contracts depends on two factors: how physical reality is represented in a smart contract's CM script *and* how the CM script encodes the material agency of digital objects (e.g., how it executes a contract if conditions are met). Because CM scripts define how smart contracts translate between states in a digital reality, they must meet a number of criteria that stretch beyond representational completeness and clarity, such as security (e.g., API integration), information quality (e.g., confidentiality, integrity, and availability of data), and transparency (e.g., data lineage and explainability of the outcomes).

### *Effecting Physical Change Through Augmented Reality*

A third key idea of our framework is that CM scripts can be tools that influence how change is effected in the physical world. Because digital objects feature in many of our everyday experiences, it is likely that the CM scripts that specify the semantics of these digital objects also influence our behaviors when we use them.

This idea presents an opportunity for CM scholarship to explore a variety of dependent variables that are new to CM. For example, research could measure *possible changes in physical reality* that stem from how physical reality is represented in CM scripts. Such a setting opens CM scholarship up to considering outcome variables that are of interest to other IS communities, such as improved decision-making, effective use, and addiction. In turn, addressing other IS communities' interests presents an opportunity for CM scholarship to build stronger conceptual bridges to other streams of IS scholarship.

To illustrate a concrete research opportunity in this area, consider wearable devices for self-management of one's health. These devices operate on scripts that capture physical bio-data like steps, pulse, and heart rate. They are designed to empower changes in human behavior that are measurable through variables like daily steps, calorie intake, and sleep score (Nelson et al. 2016).

Our framework suggests that the degree to which a healthier lifestyle is possible through wearable devices depends on the quality of the CM script that specifies the device's deep, surface, and physical structures. The deep structure semantics of devices like Fitbit are incomplete. They are limited to the physical structure elements that are available to capture biodata, such as pedometers and heart rate monitors. Other relevant data such as nutrition information is missing. This information is relevant to managing eating habits and dietary conditions but relies mainly on input in aggregated form (sum of calories) through an API that allows the device to communicate with other technologies (e.g., nutrition self-tracking software like MyFitnessTracker). Depending on the type of goal related to physical change (e.g., more sleep, more healthful eating, more exercise), situations might occur in which the wearable device could encourage users to engage in contraindicated behaviors based on the information in its CM script. For example, chronic heart conditions might conflict with increased exercise levels, more sleep might be counterproductive for amnesia patients, and healthier eating might overlook diabetic requirements.

Wearable devices could be improved by examining how their CM scripts could import data that they do not receive in sufficiently detailed form from the device's physical- or surface-structure elements. Through APIs, data could be imported from other scripts. For example, the SMART App Framework (Mandl and Kohane 2009) connects third-party applications to electronic health record data. The framework provides a reliable, secure authorization protocol for a variety of app architectures, including apps that run on an end-user's device. Through such extensions, the quality of the wearable device's CM script could be improved and new functionality added (e.g., warning signals when the heart rate levels of patients with heart conditions reach certain thresholds). Such extended CM scripts could empower behavioral changes that differ qualitatively for users with certain conditions.

## Broader Implications

Broader implications of our framework for CM scholarship reach beyond the illustration of uncharted research trajectories. First, our framework suggests a reinvigorated focus on **CM design research**. The framework centers on the mediating role of CM scripts to facilitate transitions between states of reality. Traditional CM grammars were designed with a focus on representing the transition from physical reality to digital reality alone. To extend the mediating role of CM scripts (e.g., from digital to digital), researchers may need to develop new grammars and methods, such as new grammars with new constructs to present states in digital reality, and methods for tracking changes in the states between realities. Similarly, with the blurred lines between IS's deep, physical, and surface structures in a digital world, researchers may need to develop new grammars to represent parts of and relationships between the structural elements of IS in complex digital objects.

Likewise, changes to the technological landscape (e.g., the rise of big data, intelligent machines, rampant data repurposing, wearable technologies, ubiquitous computing) require revisiting traditional CM design assumptions, which may spur the development of novel CM methods, grammars, and scripts. For example, the requirements of open information environments, where controls over information production are considerably weaker than they are in traditional organizational settings, motivate the search for novel CM approaches that are adaptable, flexible, and open (Chen 2006; Parsons and Wand 2014). Our framework recognizes that the deployment of CM scripts as mediating objects could result in an update of CM grammars or methods or changes in CM agents' views and beliefs.

A second broad implication of our framework for CM scholarship research concerns its call for research on **dependent variables** that are new to CM. The key dependent variables of quality and understandability were traditionally evaluated based on the scripts' underlying grammars. Future research should evaluate new dependent variables like the quality, consistency, traceability, and understandability of newly developed scripts and grammars. For example, with the rise of big data, open information environments, and machine learning algorithms, variables like cognitive sufficiency might be relevant to examinations of how CM scripts could be used to decrease cognitive load while increasing understanding of complex algorithms.

A third broad implication of the CMDW framework concerns the need to consider **empirical methodologies** that have not traditionally been used in CM scholarship. Our framework introduces the notion of a CM agent, which can be a human or a digital entity, who/that produces and/or consumes CM scripts. Increasing dependency on automated digital agents will increase the complexity of the nature of human–machine interactions during CM. Such a development may require CM research to adopt new modes of data collection and/or analysis (e.g., ideas from computational social science) in order to determine how digital and human agents interact during CM, based on the trace data they leave behind.

Finally, as our framework extends the scope of CM research and application, it paves the way for **collaborations between CM scholarship and other research communities**. For example, by requiring blurred boundaries between surface, physical, and deep structures to be modeled, the CMDW framework implies the need to work with research communities that focus on key elements of digital technologies surface or physical structures, such as human–computer interactions or electrical and industrial engineering. Other IS communities have studied many dependent variables that may now be relevant to CM research, so they could collaborate more actively with CM researchers in areas like effective use, machine learning accuracy, transparency and performance, and organizational change. Some recent CM research has featured such collaborations (Lukyanenko, Castellanos, et al. 2019; Nalchigar and Yu 2018), but they remain scarce. With our new framework, we encourage more interaction between CM research and other disciplines to build stronger cross-disciplinary ties.

## Conclusion ▰▰▰▰▰

CM has stood as a cornerstone of our discipline for a long time, but its standing and relevance have been repeatedly challenged, and never more than today.

To unfreeze the present stalemate between continuing high-quality CM research and CM's perceived declining relevance in IS research, we provide a reconceptualization of CM in light of a changing IS landscape. Our new framework illuminates new pathways to CM research that challenge our assumptions about what CM is. It can move our research efforts toward the fringes of the CM paradigm, where we can explore unknown territory, rather than confirm entrenched assumptions. Our framework draws attention to significant new opportunities for the CM community and substantially expands our view of what counts as CM research. The framework also brings CM research closer to other research communities, which provides opportunities for cross-pollination of ideas and interdisciplinary collaboration.

Following the research agenda suggested by our work will reveal that CM continues to have limits, but it will also increase our confidence in where, how, and why CM is effective and useful. We may even discover that CM holds promise that we have not foreseen.

## Acknowledgments

We are indebted to the senior editor, Jason Thatcher, the associate editor, Chuan-Hoo Tan, and three anonymous reviewers, for constructive and developmental feedback that helped us improve the paper. We thank participants at the 2017 AIS SIGSAND Symposium hosted by the University of Cincinnati for providing feedback on our work. We thank Jan DeGross for her production work on this paper and for all the contributions she has made to IS scholarship for many decades.

## *About the Authors*

**Jan Recker** is AIS fellow, Alexander von Humboldt Fellow, chaired professor of Information Systems and Systems Development at the University of Cologne, and adjunct professor at Queensland University of Technology. His research focuses on systems analysis and design, digital innovation and entrepreneurship, and digital solutions for sustainability challenges.

**Roman Lukyanenko** is an associate professor in the Department of Information Technologies at HEC Montréal, Canada. His research interests include conceptual modeling, information quality, crowdsourcing, machine learning, design science research, and research methods (research validities, instantiation validity, and artifact sampling). Roman's work has been published in *Nature*, *MIS Quarterly*, *Information Systems Research*, *Journal of the Association for Information Systems*, *European Journal of Information Systems*, among others. Roman is the current president for the AIS SIGSAND and the developer of www.sigsand.com.

**Mohammad Jabbari** is a lecturer in the School of Information Systems at Queensland University of Technology. He has a B.Sc. in Mathematics, a Master's degree in IT Management and a Ph.D. in Information Systems. He conducts research on systems analysis and design and conceptual modeling.

**Binny M. Samuel** is an associate professor in the Operations, Business Analytics, and Information Systems Department of the Lindner College of Business at the University of Cincinnati. He earned his Ph.D. from the Kelley School of Business at Indiana University. Prior to his doctoral education, he worked in IT roles at Ford Motor Company and at Indiana University. His work is published in *Communications of the ACM*, *European Journal of Information Systems*, *IEEE Transactions on Software Engineering*, *Journal of the Association for Information Systems*, and *MIS Quarterly*, among others.

**Arturo Castellanos** is an assistant professor in the department of information systems and statistics at the Zicklin School of Business, Baruch College (CUNY). His research interests are in the areas of system analysis and design, blockchain, and business analytics. Dr. Arturo received his M.S. and Ph.D. degrees in business (information systems) from Florida International University. He currently serves as the vice president for the AIS Special Interest Group in Systems Analysis and Design (SIGSAND). His work has been published in *Journal of the Association for Information Systems*, *Decision Support Systems*, and *Journal of Corporate Citizenship*.

# Appendix A

## Literature Review Coding Scheme ▰▰▰▰▰▰▰

| Focus/Goal of Study | What is the paper's *research objective* (Vessey et al. 2002)? |
|---|---|
| | Which CM *construct* (Wand and Weber, 2002) does the study address? (*Multiple answers possible*)<br>• Grammar • Method • Other<br>• Script • Context |
| Type of Grammar | What type of CM grammar is the focus of the paper? (*Multiple answers possible*).<br>Examples: BPMN, UML, ER. |
| Type of Study | Which *method* (Chen and Hirschheim, 2004) is used in empirical papers? (*Multiple answers possible*)<br>• Case Study • Action Research • Survey<br>• Interview • Panel • Delphi Study<br>• Experiment • Design Science • Other |
| | Which *method* (Chen and Hirschheim, 2004) is used in non-empirical papers?<br>• Literature Review • Theory Development • Commentary |
| Empirical Evidence | For empirical papers:<br>• What *empirical evidence* is provided, and what is the *sample size* (GRADE Working Group 2004)?<br>• What are *measurement items* (e.g., accuracy, time, domain understanding, perceived ease of use) and *materials* (e.g., questionnaire, interview protocol, case description, interface, developed diagrams)? |
| Assumptions | What *main assumptions* are evident in the paper (self-developed)?<br>• What perspective does a CM script *represent* (e.g., the real-world, someone's or some groups' perception of reality, a digital reality)?<br>• Which *structural aspects* of an IS are the focus of representation (e.g., deep structure, surface structure, physical structure, multiple)?<br>• What is the *purpose* and *intended outcome* of CM (e.g., CM scripts are developed for requirement analysis or to facilitate communication in organizational settings)?<br>• Who are the *actors* participating in CM production or consumption, such as human actors (in roles such as analysts, domain experts, end users) and digital actors (such as algorithms, execution engines)?<br>• What is the *context* in which CM occurs (e.g., within organizations, for professional purposes, for teaching and learning, in private life settings)? |

# Appendix B

## Summative Findings from Literature Review

Here we summarize descriptive observations from our literature review. First, more than 73 percent of the sampled articles were clearly informed and guided by the research framework Wand and Weber (2002) proposed. All of Wand and Weber's constructs received attention (Table B1): 44 papers presented research on grammars or grammars together with other constructs (e.g., grammar and method, grammar and context), 78 papers addressed methods or methods together with other constructs, 42 were on scripts or scripts together with other constructs, and 47 focused on context or context together with other constructs. Overall, 59 out of 197 papers focused on more than one CM construct.

| Table B1.  Research on Conceptual Modeling Based on the Focus of the Study | Papers | |
|---|---|---|
| **CM Construct in Focus** | **#** | **%** |
| Grammar | 20 | 10.15 |
| Method | 41 | 20.81 |
| Script | 11 | 5.58 |
| Context | 14 | 7.11 |
| Grammar and Method | 8 | 4.06 |
| Grammar and Script | 7 | 3.55 |
| Grammar and Context | 4 | 2.03 |
| Method and Script | 10 | 5.08 |
| Method and Context | 16 | 8.12 |
| Script and Context | 7 | 3.55 |
| Grammar, Method, and Script | 1 | 0.51 |
| Method, Script, and Context | 2 | 1.02 |
| Grammar, Script, and Context | 4 | 2.03 |
| Constructs other than those highlighted by Wand and Weber (2002). | 52 | 26.40 |
| **Total** | **197** | 100.00 |

Second, the grammars investigated most often in the reviewed papers were UML (70) and ER (35) (Table B2), followed by process modeling grammars, including Petri nets (13), BPMN (12) and EPC (8). This situation might indicate a stronger focus in CM research on substance and form than on behavior and change (Burton-Jones and Weber 2014).

| Table B2.  Number of Papers per Type of Grammar | | | |
|---|---|---|---|
| **Grammar** | **#** | **Grammar** | **#** |
| UML | 38 | ER | 28 |
| Specific UML grammar | 32 | Petri nets | 13 |
| • Class | 15 | BPMN | 12 |
| • Use Case | 7 | EPC | 8 |
| • Activity | 7 | Extended ER | 7 |
| • State Machine | 5 | Workflow | 4 |
| • Sequence | 4 | ANSI Flowchart | 3 |
| • Collaboration | 2 | DFD | 2 |
| • Profile | 2 | YAWL | 2 |
| MibML, ISO TC87, Merise, ebXML, BPML, WSCL, WS-BPEL, DEMO, ProH, REA, ORM, IFO, FDM, SDM, NIAM, OMT, OML | | | 1 |

Third, most of the reviewed papers were consistent with the assumptions Wand and Weber (2002) articulated (Table B3). Fourth, experiments and theoretical approaches were the research methods used most often in CM papers. Design science and action research papers are scarce (Table B4).

| Table B3. Distribution of Dominant Assumptions in CM Papers | | |
|---|---|---|
| **Representation of Reality** | **#** | **%** |
| Physical Reality | 187 | 94.9 |
| Physical and Digital Reality | 10 | 5.1 |
| Digital Reality | 0 | 0.0 |
| **IS Structure Focus** | **#** | **%** |
| Deep | 176 | 89.3 |
| Surface | 2 | 1.0 |
| Physical | 3 | 1.5 |
| Deep and Surface | 15 | 7.6 |
| Physical and Surface | 0 | 0.0 |
| Deep and Physical | 1 | 0.5 |
| Deep, Surface, and Physical | 0 | 0.0 |
| **Actors Involved in CM** | **#** | **%** |
| Human: Professional IS analysts (experienced, novice, and student proxies) | 172 | 87.3 |
| Human: End users | 2 | 1.0 |
| Human: Students (in learning and teaching) | 2 | 1.0 |
| Digital: Algorithms (script creation, script validation, script transformation, script analysis) | 18 | 9.1 |
| None explicitly identified or evident | 3 | 1.5 |
| **Purposes and Intended Outcomes of CM** | **#** | **%** |
| Domain understanding (including methods for improving understanding or model quality, etc.) | 82 | 41.6 |
| New methods (e.g., patterns, approaches, guidelines) | 21 | 10.7 |
| Automation of script creation | 17 | 8.6 |
| Evaluation of grammars and methods | 15 | 7.6 |
| Script consistency | 8 | 4.1 |
| Script validation | 7 | 3.6 |
| Script mapping | 5 | 2.5 |
| Script transformation | 5 | 2.5 |
| Script reverse engineering | 2 | 1.0 |
| Ontological foundations and evaluation of ontologies | 5 | 2.5 |
| CM tool support | 3 | 1.5 |
| CM security aspects | 2 | 1.0 |
| Other (e.g., traceability, software maintenance, usage, usefulness) | 25 | 12.7 |
| **CM Context** | **#** | **%** |
| Organizational, professional setting | 192 | 97.5 |
| Learning and teaching | 2 | 1.0 |

| Table B4. Number of Papers by CM Construct and Research Method | | | |
|---|---|---|---|
| **Focal CM Construct** | **References to Coded Papers** | **Research Method Used** | **#** |
| Grammar | Allen and March 2006, 2012; Azevedo et al. 2015; Barbier et al. 2003; Bera et al. 2014; Bera et al. 2010; Bowen et al. 2006, 2009; Briand et al. 2004; Burton-Jones et al. 2009; Chen and Carlis 2003; Clarke et al. 2016; Dobing and Parsons 2008; Dussart et al. 2004; Evermann and Fang 2010; Evermann and Wand 2009; Figl et al. 2013; Green and Rosemann 2004; Harzallah et al. 2012; Irwin and Turk 2005; La-Ongsri and Roddick 2015; Laurier and Poels 2012; Liu et al. 2004; Milicev 2007; Milton and Kazmierczak 2004; Milton et al. 2012; Opdahl and Henderson-Sellers 2004; Parsons 2011; Recker 2010; Recker et al. 2010; Recker et al. 2011; Recker et al. 2009; Rittgen 2006; Rosemann and Van der Aalst 2007; Santos et al. 2013; Shanks et al. 2010; Shanks et al. 2008; Shanks and Weber 2012; Soffer and Kaner 2011; Soffer et al. 2010; VanderMeer and Dutta 2009; Wagner 2003; Zhang et al. 2007; zur Muehlen and Indulska 2010 | Case Study | 2 |
| | | Survey | 3 |
| | | Experiment | 14 |
| | | Interview | 2 |
| | | Prototype and Formal | 4 |
| | | Commentary | 3 |
| | | Theoretical | 17 |
| Method | Ågerfalk and Eriksson 2004; Allen and March 2003; An et al. 2010; Andrade et al. 2004; Athenikos and Song 2013; Atkinson et al. 2015; Autili et al. 2015; Balaban and Shoval 2002; Basin et al. 2014; Batra 2005, 2012; Bendraou et al. 2010; Bera 2012; Bera et al. 2010; Beydoun et al. 2014; Burgueno et al. 2014; Burton-Jones and Meso 2006; Cabot et al. 2010; Chen and Carlis 2003; Chua et al. 2002; Clarke et al. 2016; Clegg and Shaw 2008; Cuadrado et al. 2014; Currim and Ram 2012; Currim et al. 2014; Cysneiros and do Prado Leite 2004; de Brock 2016; De Lara et al. 2013; Di Pietro et al. 2011; Dietz and Juhrisch 2012; Distefano et al. 2010; Domínguez et al. 2002; Dori et al. 2008; Dunn et al. 2011; Eriksson and Ågerfalk 2010; Escalona and Aragón 2008; Evermann 2005; Evermann and Wand 2005; France et al. 2004; Gómez et al. 2009; Hadar and Soffer 2006; Harzallah et al. 2012; Hsu et al. 2003; Joosten and Purao 2002; La-Ongsri and Roddick 2015; Lallchandani and Mall 2011; Lechtenbörger and Vossen 2003; Lee and Wyner 2003; Lohmann 2013; Loucopoulos and Kadir 2008; Lukyanenko et al. 2014; Ma 2005; Malavolta et al. 2009; Opdahl and Henderson-Sellers 2004; Pardillo et al. 2011; Parsons and Wand 2008, 2013; Pickin et al. 2007; Poels et al. 2011; Purao et al. 2003; Recker 2013; Reinhartz-Berger and Sturm 2008; Rittgen 2006; Rosemann and Green 2002; Rosemann and Van der Aalst 2007; Sadiq et al. 2005; Saraiva and da Silva 2008; Siau and Rossi 2011; Soffer and Hadar 2007; Soffer et al. 2015; Storey et al. 2002; Teruel et al. 2003; Vara et al. 2014; Verdickt et al. 2005; Wagner 2003; Xu and Nygard 2006; Zhao and Ram 2005 | Case Study | 11 |
| | | Action Research | 1 |
| | | Experiment | 20 |
| | | Interview | 2 |
| | | Design Science | 3 |
| | | Prototype and Formal | 7 |
| | | Theoretical | 27 |
| Script | Allen and March 2006, 2012; Autili et al. 2015; Bandara et al. 2005; Bera 2012; Bera et al. 2014; Bera et al. 2010; Beydoun et al. 2014; Bowen et al. 2006, 2009; Burton-Jones and Meso 2006, 2008; Calì et al. 2012; Chan et al. 2014; Chavez et al. 2015; Dunn et al. 2011; Egyed 2010; Evermann 2008; Figl et al. 2013; Gemino and Parker 2009; Genero et al. 2011; Halpin 2002; Khatri et al. 2006; Kimelman et al. 2009; Krogstie et al. 2006; Masri et al. 2008; Milton et al. 2012; Parsons 2002, 2011; Parsons and Wand 2008, 2013; Pickin et al. 2007; Poels et al. 2011; Recker 2013; Reder and Egyed 2013; Reinhartz-Berger and Sturm 2008; Shanks et al. 2010; Shanks et al. 2008; Siau and Lee 2004; Soffer and Kaner 2011; Sun et al. 2006 | Case Study | 1 |
| | | Experiment | 24 |
| | | Design Science | 1 |
| | | Leveraging previous data | 1 |
| | | Literature Review | 2 |
| | | Theoretical | 7 |
| Context | Allen and March 2006; Autili et al. 2015; Bandara et al. 2005; Basin et al. 2014; Bera et al. 2014; Browne and Parsons 2012; Burgueno et al. 2014; Chavez et al. 2015; Cuadrado et al. 2014; Currim et al. 2014; Damas et al. 2005; Davern et al. 2012a, 2012b; De Lara et al. 2013; Dzidek et al. 2008; Egyed 2010; Eshuis and Wieringa 2004; Evermann 2005; Figl et al. 2013; France et al. 2004; Green and Rosemann 2004; Grundy et al. 2012; Hadar and Soffer 2006; Khatri et al. 2006; Kimelman et al. 2009; Koschmider et al. 2010; Lallchandani and Mall 2011; Larsen et al. 2009; Loucopoulos and Kadir 2008; Masri et al. 2008; Pardillo et al. 2011; Pickin et al. 2007; Purao et al. 2003; Recker 2010; Recker et al. 2010; Recker et al. 2012; Reder and Egyed 2013; Sadiq et al. 2005; Samuel et al. 2015; Shanks et al. 2010; Soffer and Hadar 2007; Soffer and Wand 2007; Storey et al. 2002; Topi and Ramesh 2002; VanderMeer and Dutta 2009; Vara et al. 2014; Weidlich and Mendling 2012; Zhang et al. 2011 | Case Study | 4 |
| | | Survey | 1 |
| | | Experiment | 18 |
| | | Interview | 5 |
| | | Design Science | 1 |
| | | Prototype and Formal | 8 |
| | | Literature Review | 4 |
| | | Commentary | 2 |
| | | Theoretical | 7 |

| Focal CM Construct | References to Coded Papers | Research Method Used | # |
|---|---|---|---|
| Other | Abelló et al. 2006; Alhajj 2003; Analyti et al. 2007; Apvrille et al. 2004; Arisholm et al. 2006; Atkinson et al. 2009; Briand et al. 2005; Briand et al. 2006; Bryce et al. 2010; Chua et al. 2012; Concas et al. 2007; D'Aubeterre et al. 2008; da Silva et al. 2010; Dahanayake et al. 2003; de Albuquerque and Christ 2015; Dreiling et al. 2006; El-Attar et al. 2015; Fernández-Medina et al. 2007; Fonseca and Martin 2007; Garrigós et al. 2012; Goseva-Popstojanova et al. 2003; Gribaudo and Horváth 2002; Henderson-Sellers 2002; Karagiannis and Buchmann 2016; Kim et al. 2007; Kounev 2006; Kpodjedo et al. 2013; Lano and Kolahdouz-Rahimi 2014; Lau and Wang 2007; Leopold et al. 2014; Maté and Trujillo 2012; Mattsson et al. 2008; Milicev 2002; Miner 2006; Mitra et al. 2007; Montero et al. 2007; Piccioni et al. 2012; Recker et al. 2013; Reijers et al. 2011; Rodríguez et al. 2007; Shen et al. 2009; Shiu and Fong 2009; Shousha et al. 2010; Soffer 2005; Trujillo et al. 2004; Uchitel et al. 2003; Vergara et al. 2007; Vidyasankar and Vossen 2011; Wagelaar and Van Der Straeten 2007; Wand and Weber 2004; Weber 2003; Zhao et al. 2010 | Case Study | 16 |
| | | Action Research | 1 |
| | | Survey | 1 |
| | | Experiment | 7 |
| | | Panel | 1 |
| | | Design Science | 1 |
| | | Prototype and Formal | 1 |
| | | Literature Review | 1 |
| | | Commentary | 1 |
| | | Theoretical | 23 |